\documentclass[12pt,journal,draftclsnofoot,onecolumn]{IEEEtran} 
\makeatletter
\def\ps@headings{%
\def\@oddhead{\mbox{}\scriptsize\rightmark \hfil \thepage}%
\def\@evenhead{\scriptsize\thepage \hfil \leftmark\mbox{}}%
\def\@oddfoot{}%
\def\@evenfoot{}}
\makeatother
\usepackage{multirow}
\usepackage{amssymb}
\usepackage{amsmath}
\usepackage{amsfonts}
\usepackage{graphicx}
\usepackage{subfigure}
\usepackage{cite}
\usepackage{enumerate}
\usepackage{url}
\usepackage{bm}
\usepackage{clrscode}
\usepackage{subfigure}
\usepackage{amsthm}
\usepackage{color}
\usepackage{makecell}
\usepackage{multirow}

\begin{document}
\title{Generative AI on \emph{SpectrumNet}: An Open Benchmark of Multiband 3D Radio Maps}
\author{\IEEEauthorblockN{
		{Shuhang Zhang}, \IEEEmembership{Member, IEEE},
        {Shuai Jiang},
		{Wanjie Lin},
		{Zheng Fang},\\{Kangjun Liu}, {Hongliang Zhang}, \IEEEmembership{Member, IEEE},			
		{and Ke Chen}, \IEEEmembership{Member, IEEE}}\\
\thanks{S. Zhang, S. Jiang, W. Lin, Z. Fang, K. Liu, and K. Chen are with the Pengcheng Laboratory, Shenzhen, China (email: \{zhangshh01, jiangsh01, linwj01\}@pcl.ac.cn, fz.jun26th@gmail.com, \{liukj, chenk02\}@pcl.ac.cn).}
\thanks{H. Zhang is with Department of Electronics, Peking University, Beijing, China (email: hongliang.zhang@pku.edu.cn).}
}
\maketitle
\begin{abstract}	
Radio map is an efficient demonstration for visually displaying the wireless signal coverage within a certain region. It has been considered to be increasingly helpful for the future sixth generation~(6G) of wireless networks, as wireless nodes are becoming more crowded and complicated. However, the construction of high resolution radio map is very challenging due to the sparse sampling in practical systems. Generative artificial intelligence~(AI), which is capable to create synthetic data to fill in gaps in real-world measurements, is an effective technique to construct high precision radio maps. Currently, generative models for radio map construction are trained with two-dimension~(2D) single band radio maps in urban scenario, which has poor generalization in diverse terrain scenarios, spectrum bands, and heights. To tackle this problem, we provide a multiband three-dimension (3D) radio map dataset with consideration of terrain and climate information, named \emph{SpectrumNet}. It is the largest radio map dataset in terms of dimensions and scale, which contains the radio map of 3 spacial dimensions, 5 frequency bands, 11 terrain scenarios, and 3 climate scenarios. We introduce the parameters and settings for the \emph{SpectrumNet} dataset generation, and evaluate three baseline methods for radio map construction based on the \emph{SpectrumNet} dataset. Experiments show the necessity of the \emph{SpectrumNet} dataset for training models with strong generalization in spacial, frequency, and scenario domains. Future works on the \emph{SpectrumNet} dataset are also discussed, including the dataset expansion and calibration, as well as the extended studies on generative models for radio map construction based on the \emph{SpectrumNet} dataset.
\end{abstract}


\section{Introduction}

\subsection{Background of Radio Map}
With the rapid development of the sixth generation~(6G) of wireless networks, the increasingly complex wireless functions and the deployment of massive radio nodes necessitate a deeper understanding of radio information~\cite{NDPSLNDP2021}. A radio map, which is a representation of the characteristics of radio signals over a specific geographic area, such as the received signal strengths~(RSS)~\cite{RYKC2021}, transmission environment~\cite{YTAB2013}, and other factors, can be used to understand and visualize the coverage, quality of service, and data rate of radio signals in a given area. A radio map with high accuracy intuitively illustrates the global radio conditions of wireless propagations, based on which the performance of wireless networks in terms of spectral efficiency~\cite{BLDZ2019} and localization accuracy~\cite{WHDCJ2020} can be enhanced significantly.

Recently, many researchers have been motivated to explore radio map-assisted applications across various scenarios, ranging from localization and navigation to channel modeling and user identification. In~\cite{WWMZPP2020}, the authors proposed a deep Gaussian process for indoor radio map construction and user location estimation based on RSS samples from multiple access points. In~\cite{ZCLYZXS2020}, a mobile robotic platform was developed that simultaneously constructs spatial and radio maps to support location-based services in indoor environments. A radio map-based coverage-aware navigation approach for cellular-connected unmanned aerial vehicles was proposed in~\cite{ZXJZ2021}. Additionally, a localized statistical channel modeling approach based on radio maps, which is aware of the targeted propagation environment, was introduced in~\cite{NZZC2022}. In~\cite{SYFEM2024}, a radio map-based channel charting framework was proposed for radio fingerprinting without the need for reference~signals.

However, the construction of high accuracy radio map is very challenging for the two following reasons. \emph{First}, the sampling rate of radio sensors is too sparse for radio map construction. Specifically, precise radio information can only be obtained through the deployment of radio sensors. Due to the hardware cost and the deployment area limitation of the radio sensors, the spacial sampling rate of radio of a radio map is usually less than 1\% in practical systems~\cite{SHVR2023}. This means that over 99\% of the information needed for a radio map cannot be directly obtained through radio sensing techniques. \emph{Second}, the effective bandwidth of a radio map is determined by the working bandwidth of the radio sensors. In order to obtain the radio map that contains the signal information of common wireless systems, ranging from very low frequency~(VLF) band to Ka band, the deployment cost of the radio sensors can be extremely high.

To tackle the above challenges, generative artificial intelligence~(AI) has been considered as a promising technique for high accuracy radio map construction~\cite{XNCZKXMH2023}. A well trained generative model for radio map construction is capable to obtain the knowledge of propagation channel of wireless signals in all the working bands~\cite{SWD2023}. Based on the sparse sampled signals and the generative model, a high accuracy wide band radio map with high resolution can be obtained~\cite{KAADN2024}. A necessary foundation of training the high accuracy generative model for radio map construction is to build a massive radio map dataset that not only precisely describes the propagation feature of radio signals, but also covers various transmission scenarios and transmission frequency bands~\cite{YLKC2022}. In what follows, we introduce the related works in radio map dataset generation.

\subsection{Related Works and Our Contribution}~\label{Related Work Sec}
One of the most common way for radio map generation is the channel model based method. According to general rules of radio signal propagation, path loss is often modeled using variations of log-normal shadowing models, such as the 3GPP model described in~\cite{3GPP38901}. Given the location and configuration of transmitters, a radio map can be generated based on the adopted channel model. Works such as~\cite{TR2021,ZWWWWA2023,SFH2022} adopted channel model based radio map dataset generation and trained their generative models accordingly for radio map construction. Although the channel model-based approach is relatively easy to implement, it often lacks consistency with realistic scenarios for disregarding the radio wave propagation characteristic in specific environment along specific paths.

While the channel model-based method provides a rough estimation of radio propagation, the ray-tracing method offers a more precise and site-specific approach~\cite{HB2017}. This method simulates electromagnetic waves as rays radiating from the transmitter in a finely detailed space. These rays interact with objects in the environment, accounting for reflections and potentially even diffraction or transmission~\cite{LTPGZ2021}. Finally, the rays reaching the receiver are combined to calculate path loss or other channel information. Recently, emerging studies have been adopting the ray-tracing method for more precise radio map dataset generation. For instance, the Radio MapSeer proposed in~\cite{RYKC2021}, which contains 56,000 two-dimension~(2D) radio maps in urban scenario.

Although ray-tracing for radio map generation is highly precise, its computational cost is extremely high due to the complex calculations involving massive transmission rays. Recently, several commercial simulation platforms for ray-tracing simulations have been developed to facilitate efficient radio map generation, such as the \emph{WinProp} and the \emph{Wireless Insite}. Many researchers use these platforms to create their own datasets for training generative models for radio map construction~\cite{CLZZGY2023,HWJ2017,LZLLXXMZQX2024,JCD2024}. However, these datasets are either not publicly available~\cite{CLZZGY2023,HWJ2017}, or focus on a specific field~\cite{LZLLXXMZQX2024,JCD2024}. For example, the dataset in~\cite{LZLLXXMZQX2024} contains 2D radio maps for urban scenarios over five frequencies. The dataset in~\cite{JCD2024} contains 2D radio maps in urban scenario with directional antenna. To ensure fair and accurate evaluation of generative models for radio map construction, the development of a comprehensive benchmark radio map dataset that encompasses a wide range of scenarios and fields is essential.

Several significant gaps exist between the current datasets and the targeted benchmark dataset. \emph{First}, all existing radio map datasets focus on urban scenarios, neglecting other terrain scenarios, such as rural and mountainous area. This limitation weakens the generative models' generalization for radio map construction across different terrain scenarios. \emph{Second}, current radio map datasets only consider the electromagnetic impact of objects above the ground, such as buildings and trees, while omitting terrain information such as geographic altitude and terrain material, which significantly affects radio signal propagation. \emph{Third}, existing datasets focus solely on radio maps of terrestrial networks, and no spatial three-dimension (3D) radio map dataset containing information at different heights has been released. This omission restricts the study of generative models for radio map construction in non-terrestrial networks, such as space-air-ground integrated networks (SAGINs). \emph{Finally}, existing radio map datasets primarily focus on C-band propagation, preventing the generative model from performing radio map construction in the frequency domain.

To address the above challenges, we present the largest and most comprehensive open-source radio map dataset, named \emph{SpectrumNet}, to support the study of generative models for radio map construction. Additionally, we provide examples of experiments based on this dataset. The key contributions of this paper are summarized as follows.

\begin{enumerate}
\item We present \emph{SpectrumNet}, the largest open-source radio map dataset, containing over 300,000 radio map images. This dataset includes eleven different geographical scenarios and three climate scenarios, making it the first to consider terrain, weather, and object material information. Additionally, \emph{SpectrumNet} offers variations in five different frequencies and is the first spatial 3D dataset to include radio maps at three different height levels.
\item We introduce the open-source \emph{SpectrumNet} dataset and perform experiments to train generative models for radio map construction based on this dataset. The structural similarity~(SSIM), peak signal-to-noise ratio~(PSNR), and root mean square error~(RMSE) of three baseline methods for radio map construction are evaluated across different scenarios, frequencies, and heights.
\item Future works on the \emph{SpectrumNet} dataset, including the data expansion and the quality enhancement, are discussed. Additionally, emerging research directions of generative models for radio map construction, driven by the \emph{SpectrumNet} dataset, are envisioned.
\end{enumerate}

The rest of this paper is organized as follows. In Section~\ref{DatasetSec}, we introduce the \emph{SpectrumNet} dataset and explain its properties. In Section~\ref{ApplicationSec}, we evaluate the performance of three methods for radio map construction based on the proposed \emph{SpectrumNet} dataset in different domains. Future works for dataset enhancement and generative model studies are discussed in Section~\ref{DiscussionSec}. Finally, the conclusions are drawn in Section~\ref{Conclusions}.

\section{Dataset of \emph{SpectrumNet}} \label{DatasetSec}
In this section, we introduce the novel \emph{SpectrumNet} dataset by first comparing it to existed radio map dataset, and then describe its properties in detail.
\subsection{Dataset Description}
In this part, we compare the \emph{SpectrumNet} dataset with existed radio map datasets in~Table~\ref{Dataset Comparison}. Three of the recent released datasets introduced in Section~\ref{Related Work Sec} are listed for comparison, i.e., the \emph{Radio MapSeer} proposed in~\cite{RYKC2021}, the \emph{RadioGAT} proposed in~\cite{LZLLXXMZQX2024}, and the \emph{RMDirectionalBerlin} proposed in~\cite{JCD2024}.

\begin{table}[!ht]
\centering
\caption{Radio Map Dataset Comparison}\label{Dataset Comparison}
\begin{tabular}{|c|c|c|c|c|}
\hline
\textbf{Properties} & \textbf{SpectrumNet} & \textbf{Radio MapSeer}~\cite{RYKC2021} & \textbf{RadioGAT}~\cite{LZLLXXMZQX2024} & \textbf{RMDirectionalBerlin}~\cite{JCD2024}\\
\hline
\textbf{Number of image samples} & 300k & 56k & 21k & 75k \\
\hline
\textbf{Spacial environmental dimension} & 3D & 2D & 2D & 3D \\
\hline
\textbf{Spacial data dimension} & 3D & 2D & 2D & 2D \\
\hline
\textbf{Scenario types} & 11 scenarios & Urban & Urban & Urban \\
\hline
\textbf{Terrain information} & Yes & No & No & No \\
\hline
\textbf{Climate impact} & Yes & No & No & No \\
\hline
\textbf{Frequency diversity} & 5 & 1 & 5 & 1 \\
\hline
\textbf{Number of transmitters} & Multiple & Single & Multiple & Single \\
\hline
\end{tabular}
\end{table}

\textbf{Number of image samples:} The proposed \emph{SpectrumNet} dataset is over 4 times larger than any other released radio map dataset in terms of the number of radio maps, with a total of over 300,000 images. The number of image samples will be further increased in future versions.

\textbf{Spacial dimension:} \emph{SpectrumNet} dataset is the only one that contains spacial 3D radio maps, which can be utilised to train generative models for radio map construction in the critical SAGINs in the 6G of wireless networks. Although some released datasets, such as~\cite{JCD2024} and the extended version of \emph{Radio MapSeer}~\cite{YLKC2022}, consider 3D building models for radio map construction, their samples are still 2D images that only show radio propagation in terrestrial networks.

\textbf{Scenario types:} Existed radio map datasets primarily focus on the urban scenario with massive buildings, resulting in a weak generalization capacity of generative models for radio map construction across different terrain scenarios. To address this problem, \emph{SpectrumNet} dataset includes 11 terrain scenarios with different geographical characteristics: dense urban, ordinary urban, rural, suburban, mountainous, forest, desert, grassland, island, ocean, and lake. Each type of terrain scenario features different radio propagation parameters due to the varying materials and distribution of reflectors and refractors.

\textbf{Terrain information:} Existing radio map datasets only consider the reflection and refraction effects of objects on the ground, such as buildings and trees. However, terrain information, which significantly impacts radio propagation, especially in mountainous and island areas, has not been considered. In the \emph{SpectrumNet} dataset, we account for the joint impact of terrain and ground objects on radio propagation, providing a model that better suits real-world channel conditions.

\textbf{Climate impact:} Climate information, such as temperature and rain rate, has visible impact on the radio propagation. To demonstrate this impact, the \emph{SpectrumNet} dataset includes parameters for air pressure, water density, temperature, and rain rate during radio map generation. Additionally, three different climate types: tropics, subtropics, and temperate zones,\footnote{The frigid zone is not included due to less human activities.} are included with different distributions of climate parameters.

\textbf{Frequency diversity:} To enhance the frequency generalization capacity of generative models for radio map construction, the \emph{SpectrumNet} dataset includes five different frequency bands, which is the same as dataset \emph{RadioGAT}. These are the only two open-source datasets that support frequency domain inference for radio map constructions.


\textbf{Number of transmitters:} Training generative models for radio map construction with multiple transmitters is more complicated than with a single transmitter due to the superposition effects of radio signals. Therefore, it is essential to provide samples with multiple transmitters for constructing radio maps in complex environments. \emph{SpectrumNet} and \emph{RadioGAT} are the only two datasets that include multiple transmitters in their radio maps.

\subsection{Dataset Properties}
In this part, we introduce the properties of the \emph{SpectrumNet} dataset. Table~\ref{Dataset Properties} shows the parameters of the samples in the \emph{SpectrumNet} dataset. OpenStreetMap~\cite{OSM} is used to obtain real-world building maps of 15300 different areas. The size of each area is $1.28 km \times 1.28 km$ in the x-y plane, with a spacial resolution of $10 m$, resulting in radio map images of size $128\times 128$. The radio maps are provided at three different heights for each area: $1.5 m$, $30 m$, and $200 m$ above the ground, reflecting the radio maps for ground nodes, nodes in buildings, and aerial nodes, respectively. In order to illustrate full band radio propagation, radio maps of five different frequencies ranging from 150 MHz to 22GHz are included, which cover most of the working frequencies of wireless networks. Additionally, the variation of scenarios corresponds to different climates, building materials, and terrain materials, which also affect radio propagation in real world. Therefore, these parameters are also annotated for the generation of every radio map. The open-source \emph{SpectrumNet} dataset is available at \emph{https://github.com/ShuhangZhang/FDRadiomap}.\footnote{The dataset will be uploaded after paper publication.} The documentation of the \emph{SpectrumNet} dataset is provided in Appendix~\ref{Appendix1}.

\begin{table}[!ht]
\centering
\caption{Properties of the \emph{SpectrumNet} Dataset}\label{Dataset Properties}
\begin{tabular}{|c|c|}
\hline
\textbf{Properties} & \textbf{SpectrumNet} \\
\hline
\textbf{Real world maps} & 15300 \\
\hline
\textbf{Projection area of each map} & $1.28 km \times 1.28 km$ \\
\hline
\textbf{Spatial resolution} & $10 m$ \\
\hline
\textbf{Scenario types} & \makecell[c]{Dense urban, ordinary urban, rural, suburban, mountainous, \\forest, desert, grassland, island, ocean, lake} \\
\hline
\textbf{Sample heights} & $1.5 m, 30 m, 200 m$ \\
\hline
\textbf{Frequencies} & 150 MHz, 1.5 GHz, 1.7 GHz, 3.5 GHz, 22 GHz \\
\hline
\textbf{Climate types} & Tropics, subtropics, temperate zone \\
\hline
\textbf{Building materials} & Cement, marble, glass, plywood, wood \\
\hline
\textbf{Terrain material} & Soil, grassland, water surface, cement \\
\hline
\end{tabular}
\end{table}



\subsubsection{Terrain scenarios in SpectrumNet dataset}

\begin{figure}[!ht]
\centering
\includegraphics[width=6.5in]{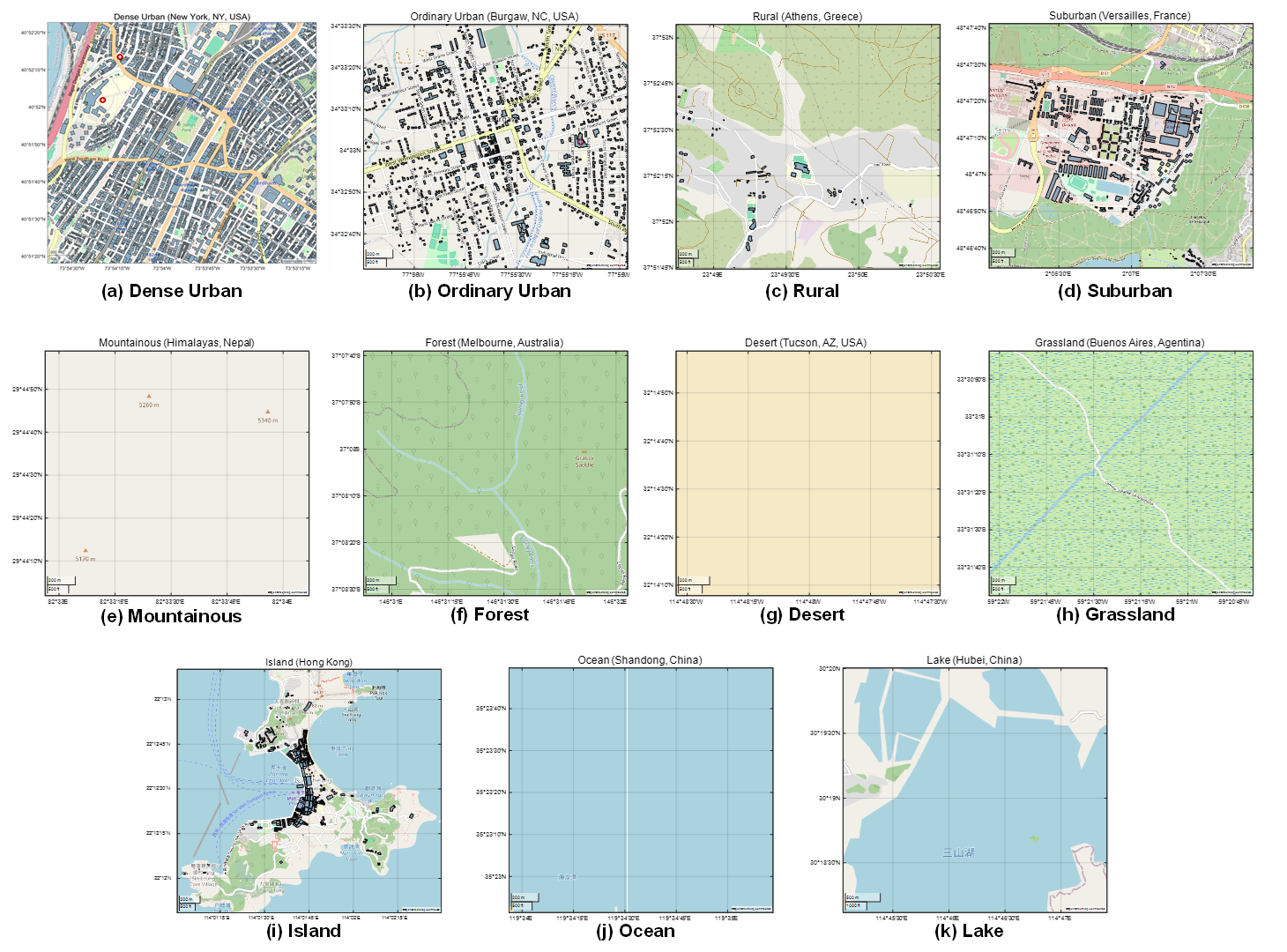}
\caption{Terrain scenarios in \emph{SpectrumNet} dataset.}
\label{Scenario}
\end{figure}

Fig.~\ref{Scenario} presents examples of the 11 different terrain scenarios in the \emph{SpectrumNet} dataset. As mentioned above, different scenarios have various parameters in terms of terrain materials, object materials, and geographical characteristics. In what follows, we briefly introduce the features of each terrain scenario in \emph{SpectrumNet}.

\begin{enumerate}[(1)]
\item \textbf{Dense urban:} Dense urban areas are characterized by a high density of tall buildings and streets, such as downtown New York City shown in Fig.~\ref{Scenario}~(a). These areas feature a significant number of rays reflecting and refracting off surfaces. The material of buildings can be cement, marble, glass, and plywood, while the terrain material is mostly cement.
\item \textbf{Ordinary urban:} Ordinary urban areas have a lower density of buildings, and the buildings are shorter compared to those in dense urban areas. Small cities and the non-central areas of big cities are typical examples of ordinary urban area, such as Burgaw, North Carolina, shown in Fig.~\ref{Scenario}~(b). Both the building material and terrain material in these areas are predominantly cement.
\item \textbf{Rural:} Rural areas feature a low density of low-height buildings and large amounts of open space, often utilized as farmland. The terrain material in these areas is predominantly soil. An example of a rural area is near Athens, Greece, as shown in Fig.~\ref{Scenario}~(c).
\item \textbf{Suburban:} In suburban areas, buildings are often clustered within specific zones, with very few buildings outside these areas. The building material is typically cement, while the terrain material includes both soil and cement. The suburban area of Versailles, France, shown in Fig.~\ref{Scenario}~(d), is a typical example of a suburban area.
\item \textbf{Mountainous:} A mountainous area is characterized by significant altitude changes and the presence of large landforms that rise prominently above their surroundings, such as the Himalayas in Nepal, shown in Fig.~\ref{Scenario}~(e). Radio propagation in mountainous areas is primarily determined by terrain information, where the material is considered as soil.
\item \textbf{Forest:} Forest areas typically have flat terrain with grassland material and no buildings. The majority of objects in forests are trees made of wood. An example of a forest area is located near Melbourne, Australia, as shown in Fig.~\ref{Scenario}~(f).
\item \textbf{Desert:} Desert areas are an open spaces with flat terrain, and have no buildings or other objects on the ground. The terrain material is set as soil. An example of desert area is located near Tucson, Arizona, as shown in Fig.~\ref{Scenario}~(g).
\item \textbf{Grassland:} Similar to desert areas, grassland areas are also open spaces with flat terrain and no buildings or other objects on the ground. The key difference is that the terrain material is grassland, which has different reflection and refraction characteristics compared to soil. An example of a grassland area is near Buenos Aires, Argentina, as shown in Fig.~\ref{Scenario}~(h).
\item \textbf{Island:} Island areas have a large proportion of water surface. The land part of an island may have significant altitude changes from the water surface. The terrain material can be cement, soil, or grassland, and the building material is typically cement. An example of an island area is located in Hong Kong, as shown in Fig.~\ref{Scenario}~(i).
\item \textbf{Ocean:} Ocean areas are completely covered by water surfaces, and have no buildings or other objects. Radio propagation in these areas involves either direct links or reflections and refractions off the water surface. An example of ocean area, as shown in Fig.~\ref{Scenario}~(j), locates in Shandong, China.
\item \textbf{Lake:} Lake areas are mostly covered by water surfaces. The main difference between lake and ocean areas is that the surrounding terrestrial terrain and buildings impact radio propagation. An example of a lake area in Hubei, China, is shown in Fig.~\ref{Scenario}~(k).
\end{enumerate}

\subsubsection{3D Terrain and Building Information in the SpectrumNet Dataset}

\begin{figure}[!ht]
\centering
\includegraphics[width=6in]{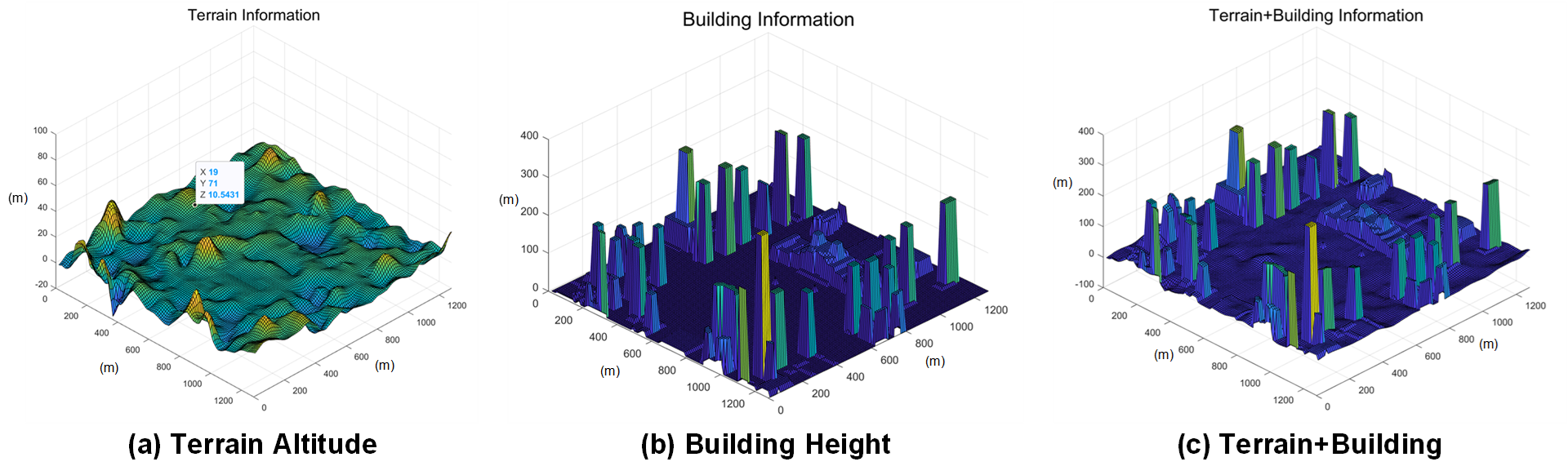}
\caption{Terrain and building information.}
\label{Terrain}
\end{figure}

The radio maps in the \emph{SpectrumNet} dataset are generated based on the 3D terrain and building information from the real world. Fig.~\ref{Terrain} illustrates the 3D terrain and building modeling process for radio map generation. We first import the 3D terrain information from OpenStreetMap~\cite{OSM}, as shown in Fig.~\ref{Terrain}~(a), which includes the altitude and material of each sampling point in the x-y plane. Next, the building information, including building height and material, is constructed based on online data from OpenStreetMap and manual supplementation using Google Maps, as shown in Fig.~\ref{Terrain}~(b). Finally, we combine the terrain and building information to create the 3D scenario model for the \emph{SpectrumNet} dataset, as illustrated in Fig.~\ref{Terrain}~(c).

\begin{figure}[!ht]
\centering
\includegraphics[width=3in]{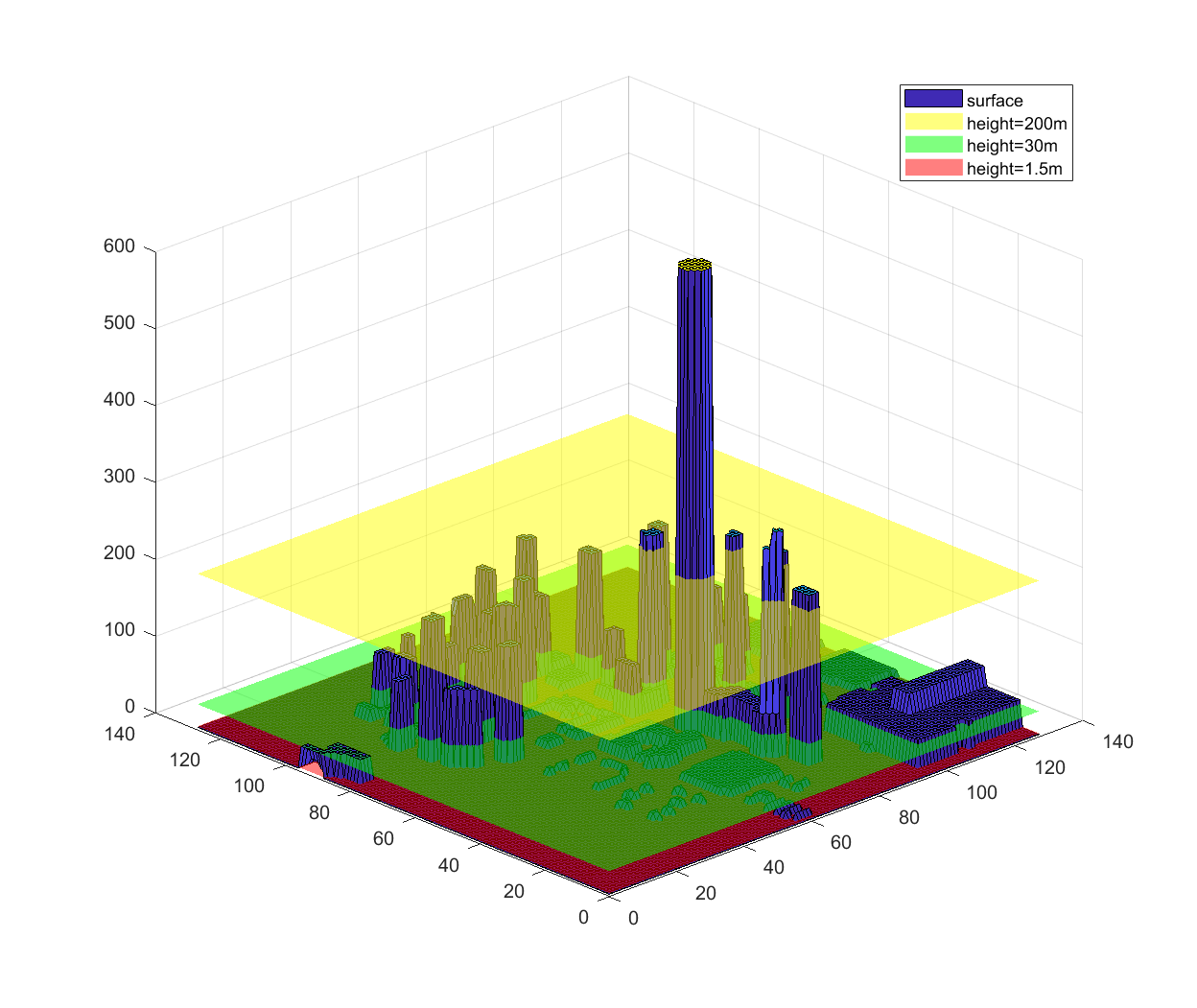}
\caption{Illustration for 3D map with terrain and building information.}
\label{3DMap}
\end{figure}

Based on the 3D scenario model, we construct the 3D radio map by sampling the RSS at different heights. The sampling layers of the spatial 3D radio map in the \emph{SpectrumNet} dataset are presented in Fig.~\ref{3DMap}. A sampling layer represents the radio map at a specific height above the terrain. \emph{SpectrumNet} contains three sampling layers at heights of $1.5 m$, $30 m$, and $200 m$, which represent the radio maps of terrestrial networks, in-building networks, and aerial networks, respectively.

\begin{figure}[!ht]
\centering
\includegraphics[width=6.5in]{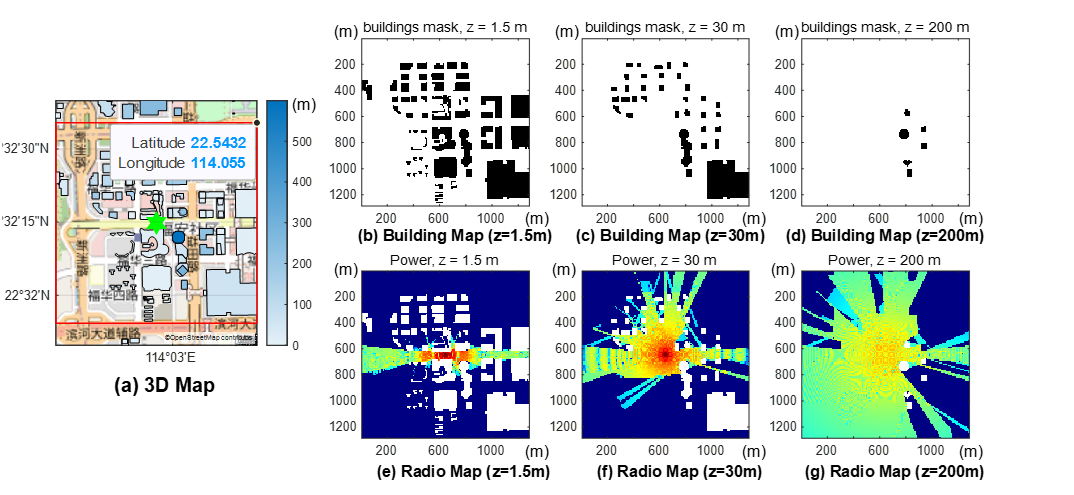}
\caption{Building maps and radio maps on different heights.}
\label{Spacial3D}
\end{figure}

Fig.~\ref{Spacial3D} provides an example of a 3D radio map with multiple sampling layers. The 3D map is built based on a dense urban scenario in Shenzhen, China, as shown in Fig.~\ref{Spacial3D}~(a), with a transmitter located at the center of this area, and the height of the transmitter is $1.5 m$. Fig.~\ref{Spacial3D}~(b), Fig.~\ref{Spacial3D}~(c), and Fig.~\ref{Spacial3D}~(d) show the projections of buildings in the sampling layers at heights of $1.5 m$, $30 m$, and $200 m$, respectively. Fig.~\ref{Spacial3D}~(e), Fig.~\ref{Spacial3D}~(f), and Fig.~\ref{Spacial3D}~(g) present the radio maps of the corresponding sampling layers. The building density of the lowest sampling layer in Fig.~\ref{Spacial3D}~(b) is higher than that of the highest sampling layer in Fig.~\ref{Spacial3D}~(d). As a result, the radio signal coverage in Fig.~\ref{Spacial3D}~(e) is less than the radio coverage in Fig.~\ref{Spacial3D}~(g) due to a more severe shadow effect. Conversely, the RSS within the coverage area in Fig.~\ref{Spacial3D}~(e) is higher than that in Fig.~\ref{Spacial3D}~(g) because of a stronger reflection effect.

Moreover, it can be observed that the shadow effect of a radio map is not solely determined by the projection of buildings in the same sampling layer. For example, in Fig.~\ref{Spacial3D}~(d), the projections of all buildings in the highest sampling layer are distributed on the right side of the transmitter, yet there is still a significant shadow effect on the left side caused by buildings of different heights. This phenomenon demonstrates the necessity of building a 3D radio map dataset rather than a 2D radio map dataset in training generative models for radio map generation in non-terrestrial networks.

\subsubsection{Frequency Variance in the SpectrumNet Dataset}

\begin{figure}[!ht]
\centering
\includegraphics[width=5in]{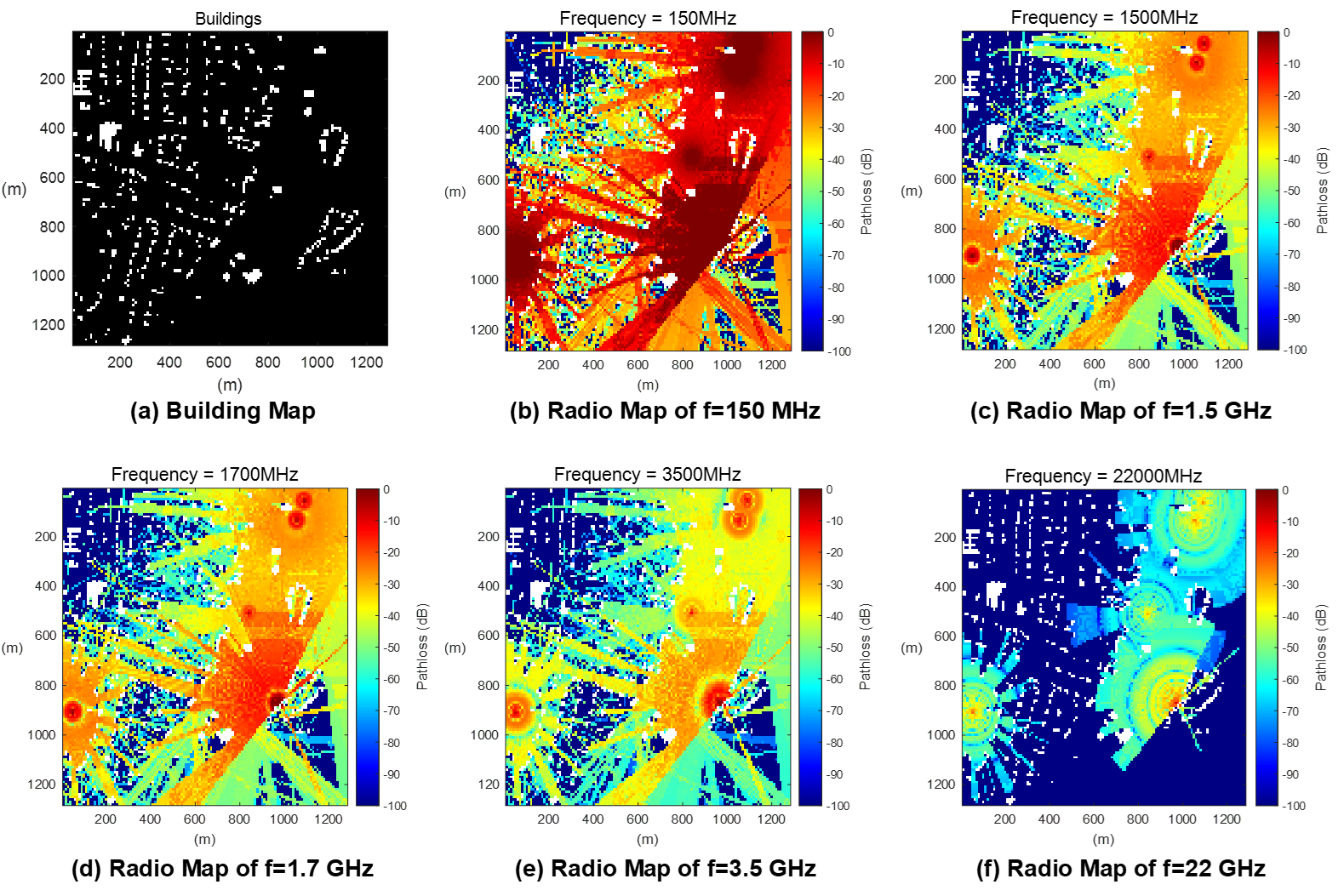}
\caption{Radio maps of different bands in frequency domain.}
\label{Frequency}
\end{figure}

The radio propagation characteristics vary significantly across different frequency bands. To improve the frequency domain generalization capacity of generative models for radio map construction, the \emph{SpectrumNet} dataset includes radio maps of five different frequency bands. Fig.~\ref{Frequency} shows a 2D projection of the buildings and radio maps at different frequency bands with a height of $1.5 m$. It can be observed that the large scale fading effect differs prominently between the 150 MHz network in Fig.~\ref{Frequency}(b) and the 22 GHz network in Fig.~\ref{Frequency}(f). This highlights the importance of including radio maps of various frequency bands in the \emph{SpectrumNet} dataset to ensure effective frequency domain generalization.

\subsubsection{Weather Parameters in the SpectrumNet Dataset}

\begin{table}[!ht]
\centering
\caption{Example of Weather Parameters}\label{Weather Parameter}
\begin{tabular}{|c|c|c|c|c|c|c|c|}
\hline
\textbf{City} & \textbf{Weather} & \textbf{Level}& \textbf{\makecell[c]{Temperature \\($^\circ C$)}}& \textbf{\makecell[c]{Air Pressure\\ (Pa)}}& \textbf{\makecell[c]{Water Vapor \\Density ($g/m^3$)}}& \textbf{\makecell[c]{Liquid Water \\Density ($g/m^3$)}}& \textbf{\makecell[c]{Rain Rate \\($mm/h$)}}\\
\hline
Jakarta & Sunny & Heavy& 36 - 40& 99800 - 100400& 50 - 100& 5 - 20& -\\
\hline
Jakarta & Sunny & Middle& 20 - 38& 101000 - 101400 & 25 - 50& 5 - 15& -\\
\hline
Jakarta & Sunny & Light& 20 - 35& 101100 - 101500 & 10 - 32& 1 - 10& -\\
\hline
Jakarta & Rainy & Heavy& 37 - 43& 100800 - 101200& 35 - 80& 5 - 20& 250 - 500\\
\hline
Jakarta & Rainy & Middle& 32 - 40& 101000 - 101400 & 25 - 100& 3 - 20& 10 - 200\\
\hline
Jakarta & Rainy & Light& 25 - 32& 101100 - 101500 & 15 - 30& 1 - 10& 25 - 100\\
\hline
\end{tabular}
\end{table}

Various weather parameters are considered in the construction of the \emph{SpectrumNet} dataset. Based on the ITU specifications~\cite{ITU1,ITU2,ITU3}, we have built weather models for over 20 cities and regions, taking into account their temperature, air pressure, water vapor density, liquid water density, rain rate, etc. Table~\ref{Weather Parameter} lists the weather parameters for Jarkata as a representative city of tropical climate. For the generation of each radio map, we first randomly select a weather condition for the specific city, such as heavy rain in Jakarta, and then generate the weather parameters within the range provided in the fourth row of Table~\ref{Weather Parameter}. Finally, we integrate the impact of these weather parameters into the ray tracing-based radio map generation as specified in~\cite{ITU1,ITU2,ITU3}.

\section{Experiments on Generative Models for Radio Map Construction}\label{ApplicationSec}
In this section, we present baseline methods for constructing radio maps using the \emph{SpectrumNet} dataset. We evaluate the performance of these methods across various terrain scenarios, heights, and frequencies, and discuss the generalization of models trained on the \emph{SpectrumNet} dataset. Three baseline methods are employed: UNet~\cite{RFB2015}, CBAM~\cite{WPLK2018}, and an interpolation algorithm~\cite{R2017}. The three baseline methods cover the main research directions of radio map constructions. UNet and CBAM represent generative model-driven radio map construction methods based on convolutional neural network and Transformer, respectively; while interpolation is a representative wireless channel-driven radio map construction method.

We randomly select 50 sampling points as input for radio map construction, with the goal of generating a $128 \times 128$ radio map, resulting in a sampling rate of 0.3\%. For the first two generative model-driven methods, the input comprises three channels: RSS, terrain altitude, and building information. The RSS and terrain altitude values are normalized to a range between 0 and 1, while the building information is represented as a binary value (1 for locations with buildings, 0 for those without). All experiments are conducted on an NVIDIA V100 GPU. For each task, the dataset is divided into training, validation, and testing sets in a 7:1:2 ratio. The performance of the models is evaluated using three metrics: SSIM, PSNR, and RMSE.

\subsection{Radio Map Construction in Various Terrain Scenarios}
In this part, we evaluate the performance of radio map generation across various terrain scenarios, with results presented in Table~\ref{ScenarioExp}. The performance metrics, i.e., SSIM, PSNR, and RMSE, generally show consistent trends: higher SSIM values typically correspond to higher PSNR and lower RMSE values. However, the effectiveness of the baseline methods varies significantly depending on the terrain scenarios. Radio maps generated for ocean scenarios, where the terrain is flat and devoid of buildings, achieve the best performance. Conversely, scenarios with complex terrain or numerous buildings and objects, such as dense urban, suburban, and forest areas, yield lower performance with the baseline methods, indicating a need for further research on generative models for radio map construction. Among the baseline methods, CBAM outperforms the others in most cases, highlighting the superiority of this generative model in radio map applications. The performance of UNet and the interpolation algorithm is comparable, with UNet slightly better in PSNR, while the interpolation algorithm performs marginally better in SSIM and RMSE.

\begin{table}[!ht]
\centering
\caption{Performance evaluation of radio map construction in various scenarios}\label{ScenarioExp}
\begin{tabular}{|c|c|c|c|c|c|c|}
\hline
\multicolumn{7}{|c|}{\textbf{SSIM}}\\
\hline
\textbf{Test Scenario} & Grassland & Island & Ocean & Lake & Suburban & Dense urban \\
\hline
\textbf{UNet} & 0.675 & 0.679 & 0.843 & 0.817 & 0.659 & 0.638 \\
\hline
\textbf{CBAM} & 0.701 & 0.701 & 0.848 & 0.831 & 0.668 & 0.603 \\
\hline
\textbf{Interpolation} & 0.698 & 0.695 & 0.836 & 0.826 & 0.683 & 0.659 \\
\hline
\textbf{Test Scenario} & Rural & Ordinary urban & Desert & Mountainous & Forest &\\
\hline
\textbf{UNet} & 0.623 & 0.623 & 0.620 & 0.651 & 0.660 &\\
\hline
\textbf{CBAM} & 0.623 & 0.633 & 0.650 & 0.681 & 0.674 &\\
\hline
\textbf{Interpolation} & 0.657 & 0.701 & 0.648 & 0.678 & 0.677 &\\
\hline
\multicolumn{7}{|c|}{\textbf{PSNR}}\\
\hline
\textbf{Test Scenario} & Grassland & Island & Ocean & Lake & Suburban & Dense urban \\
\hline
\textbf{UNet} & 20.474 & 20.393 & 29.603 & 24.508 & 19.976 & 19.874 \\
\hline
\textbf{CBAM} & 21.378 & 21.002 & 30.320 & 25.441 & 20.487 & 20.193 \\
\hline
\textbf{Interpolation} & 18.889 & 18.619 & 27.155 & 22.949 & 18.334 & 18.650 \\
\hline
\textbf{Test Scenario} & Rural & Ordinary urban & Desert & Mountainous & Forest &\\
\hline
\textbf{UNet} & 19.624 & 19.834 & 19.829 & 20.024 & 19.961 & \\
\hline
\textbf{CBAM} & 19.977 & 20.158 & 20.482 & 20.896 & 20.800 &\\
\hline
\textbf{Interpolation} & 18.079 & 18.137 & 17.971 & 18.307 & 18.315 &\\
\hline
\multicolumn{7}{|c|}{\textbf{RMSE}}\\
\hline
\textbf{Test Scenario} & Grassland & Island & Ocean & Lake & Suburban & Dense urban \\
\hline
\textbf{UNet} & 0.095 & 0.096 & 0.033 & 0.060 & 0.101 & 0.105 \\
\hline
\textbf{CBAM} & 0.086 & 0.090 & 0.031 & 0.054 & 0.095 & 0.101 \\
\hline
\textbf{Interpolation} & 0.114 & 0.118 & 0.044 & 0.072 & 0.122 & 0.120 \\
\hline
\textbf{Test Scenario} & Rural & Ordinary urban & Desert & Mountainous & Forest &\\
\hline
\textbf{UNet} & 0.106 & 0.106 & 0.102 & 0.100 & 0.101 &\\
\hline
\textbf{CBAM} & 0.102 & 0.102 & 0.095 & 0.090 & 0.091 &\\
\hline
\textbf{Interpolation} & 0.127 & 0.129 & 0.127 & 0.122 & 0.122 &\\
\hline
\end{tabular}
\end{table}

\begin{figure}[!ht]
\centering
\includegraphics[width=5.1in]{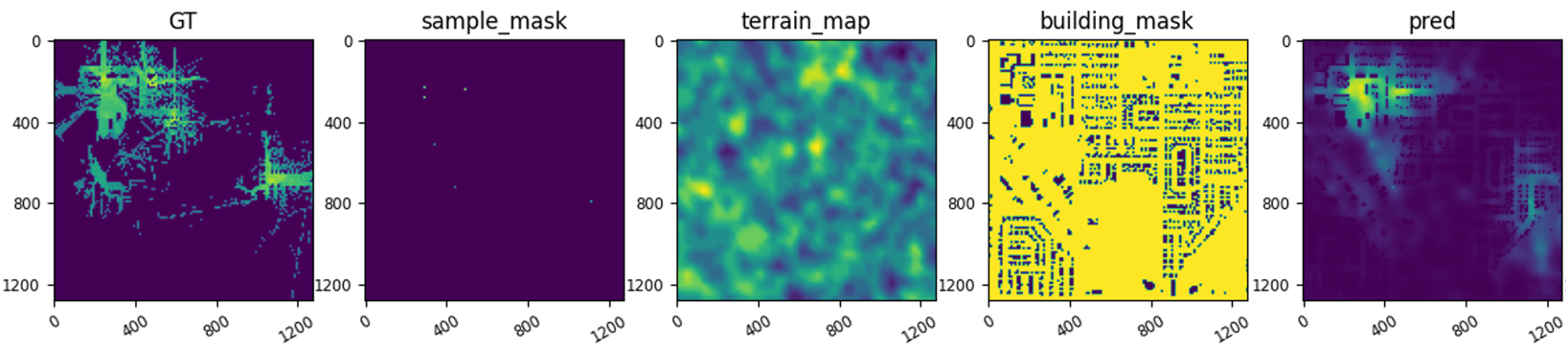}
\caption{Demonstration for radio map construction in dense urban scenario.}
\label{DenseUrbanMap}
\end{figure}

\begin{figure}[!ht]
\centering
\includegraphics[width=5.1in]{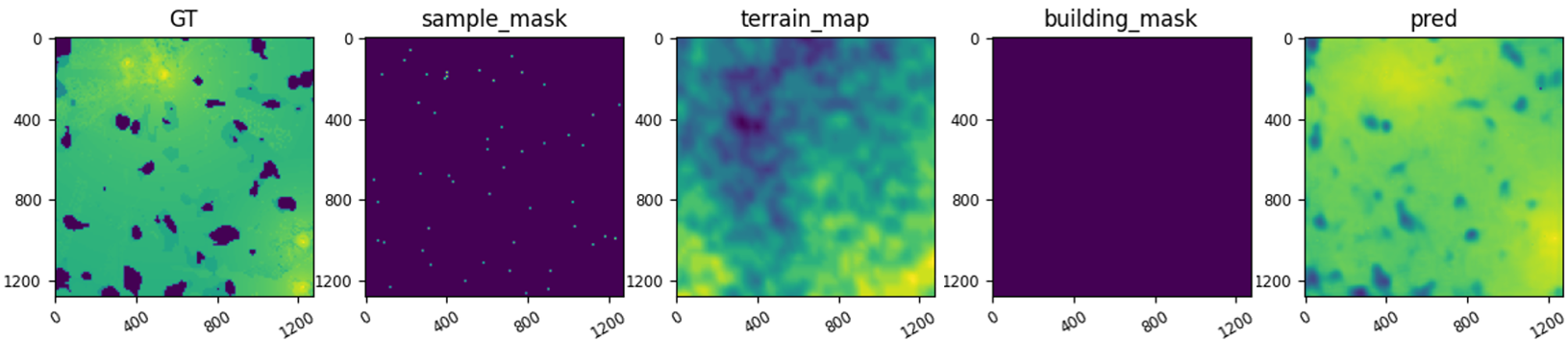}
\caption{Demonstration for radio map construction in mountainous scenario.}
\label{MountainousMap}
\end{figure}

For better illustration, we provide two examples of radio map generation in dense urban and mountainous scenarios, shown in Fig.~\ref{DenseUrbanMap} and Fig.~\ref{MountainousMap}, respectively. In each figure, the first subfigure displays the ground truth~(GT) of the radio map. The second, third, and forth subfigures show the input information of the sampling points: RSS, terrain altitudes, and building information, respectively. The fifth subfigure presents the radio map generated by the CBAM method. Fig.~\ref{DenseUrbanMap} illustrates the significant shadowing effects caused by buildings in the dense urban scenario, while Fig.~\ref{MountainousMap} highlights the dominant influence of terrain in the mountainous scenario.

\subsection{3D Radio Map Construction}
In this part, we evaluate radio map construction at different heights. Table~\ref{HeightExp} summarizes the performance of radio map generation at three distinct heights provided in the \emph{SpectrumNet} dataset. The aerial network at a height of $200 m$ achieves the best performance, while the terrestrial network at $1.5 m$ performs the worst. This disparity is attributed to the greater impact of shadowing and reflection effects from buildings and terrain on the terrestrial network, making it more complex than the aerial network. As a result, constructing accurate radio maps for terrestrial networks is more challenging compared to aerial networks.

\begin{table}[!ht]
\centering
\caption{Performance evaluation of radio map construction in different heights}\label{HeightExp}
\begin{tabular}{|c|c|c|c|}
\hline
\multicolumn{4}{|c|}{\textbf{SSIM}}\\
\hline
\textbf{Test Scenario} & 1.5 m & 30 m & 200 m \\
\hline
\textbf{UNet} & 0.556 & 0.687 & 0.820 \\
\hline
\textbf{CBAM} & 0.560 & 0.696 & 0.829 \\
\hline
\textbf{Interpolation} & 0.580 & 0.701 & 0.818 \\
\hline
\multicolumn{4}{|c|}{\textbf{PSNR}}\\
\hline
\textbf{Test Scenario} & 1.5 m & 30 m & 200 m \\
\hline
\textbf{UNet} & 18.687 & 21.661 & 26.652 \\
\hline
\textbf{CBAM} & 18.923 & 21.985 & 27.354 \\
\hline
\textbf{Interpolation} & 16.165 & 19.940 & 25.071 \\
\hline
\multicolumn{4}{|c|}{\textbf{RMSE}}\\
\hline
\textbf{Test Scenario} & 1.5 m & 30 m & 200 m \\
\hline
\textbf{UNet} & 0.120 & 0.083 & 0.047 \\
\hline
\textbf{CBAM} & 0.117 & 0.080 & 0.043 \\
\hline
\textbf{Interpolation} & 0.161 & 0.101 & 0.056 \\
\hline
\end{tabular}
\end{table}

\begin{figure}[!ht]
\centering
\includegraphics[width=5.1in]{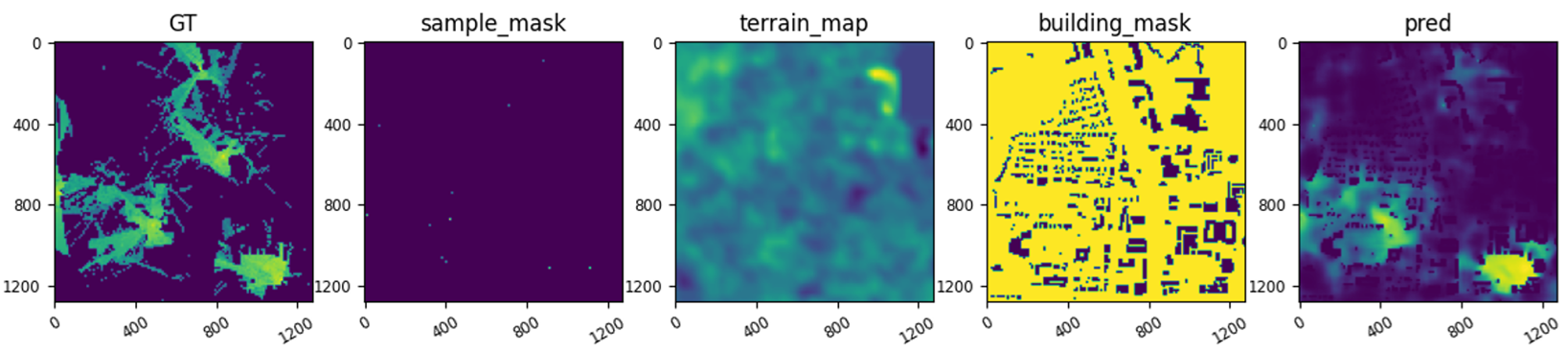}\vspace{-3mm}
\caption{Demonstration for radio map construction in terrestrial network.}\vspace{-3mm}
\label{TerrestrialMap}
\end{figure}

\begin{figure}[!ht]
\centering
\includegraphics[width=5.1in]{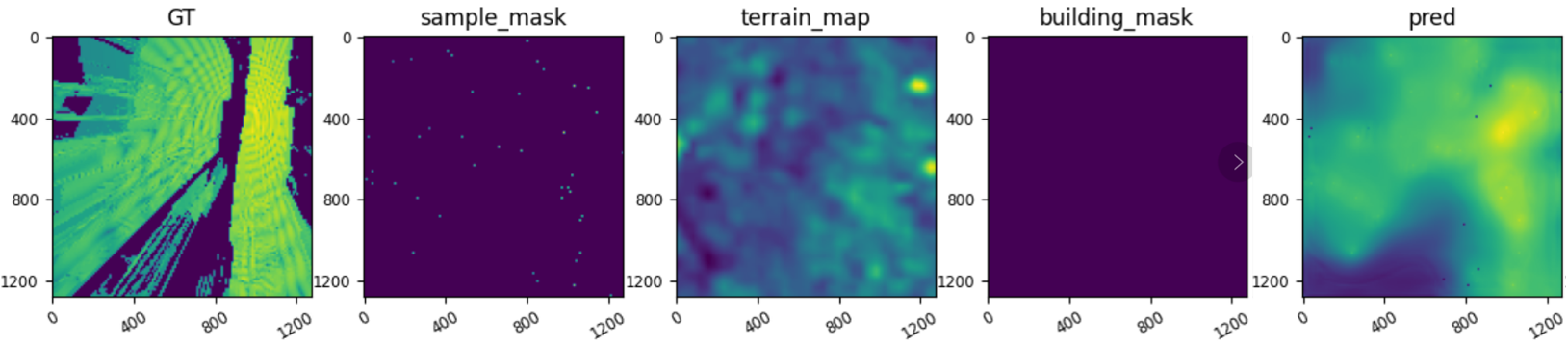}\vspace{-3mm}
\caption{Demonstration for radio map construction in aerial network.}\vspace{-3mm}
\label{AerialMap}
\end{figure}

An example of radio map generation in a terrestrial network is shown in Fig.~\ref{TerrestrialMap}. The terrestrial network radio map contains numerous detailed features that baseline methods struggle to reconstruct accurately, highlighting the need for further research in this area. Fig.~\ref{AerialMap} presents a challenging case for radio map construction in an aerial network. Although there is no building at the 200 m height in this area, the radio map still exhibits significant shadowing effects caused by buildings. This occurs because the transmitter is positioned on the ground, and the low-rise buildings that obstruct signals to the aerial network are not represented in the aerial layer. This example underscores the importance of incorporating 3D information, rather than relying solely on 2D data, into the generative model to address such challenges effectively.

\subsection{Multiband Radio Map Construction}
In this part, we evaluate radio map construction across different frequencies. Table~\ref{BandExp} presents the performance results for five distinct frequency bands. Constructing radio maps for high-frequency networks is notably more challenging than for low-frequency networks, particularly in terms of the SSIM metric. This difficulty arises from the physical characteristics of radio propagation. High-frequency signals experience more severe attenuation due to transmission and reflection by terrain and buildings, making it harder to accurately construct radio maps in environments with complex terrain and building structures. Consequently, further research is urgently needed to develop generative models capable of effectively handling radio map construction in high-frequency networks.

\begin{table}[!ht]
\centering
\caption{Performance evaluation of radio map construction in different frequency bands}\label{BandExp}
\begin{tabular}{|c|c|c|c|c|c|}
\hline
\multicolumn{6}{|c|}{\textbf{SSIM}}\\
\hline
\textbf{Test Scenario} & 150 MHz & 1.5 GHz & 1.7 GHz & 3.5 GHz & 22 GHz \\
\hline
\textbf{UNet} & 0.719 & 0.676 & 0.672 & 0.663 & 0.629 \\
\hline
\textbf{CBAM} & 0.740 & 0.694 & 0.691 & 0.667 & 0.637 \\
\hline
\textbf{Interpolation} & 0.751 & 0.699 & 0.701 & 0.684 & 0.663 \\
\hline
\multicolumn{6}{|c|}{\textbf{PSNR}}\\
\hline
\textbf{Test Scenario} & 150 MHz & 1.5 GHz & 1.7 GHz & 3.5 GHz & 22 GHz \\
\hline
\textbf{UNet} & 19.073 & 20.711 & 20.902 & 21.451 & 22.260 \\
\hline
\textbf{CBAM} & 19.522 & 21.130 & 21.350 & 21.850 & 22.529 \\
\hline
\textbf{Interpolation} & 17.391 & 19.118 & 19.330 & 19.928 & 20.094 \\
\hline
\multicolumn{6}{|c|}{\textbf{RMSE}}\\
\hline
\textbf{Test Scenario} & 150 MHz & 1.5 GHz & 1.7 GHz & 3.5 GHz & 22 GHz \\
\hline
\textbf{UNet} & 0.113 & 0.093 & 0.092 & 0.086 & 0.078 \\
\hline
\textbf{CBAM} & 0.107 & 0.089 & 0.087 & 0.082 & 0.076 \\
\hline
\textbf{Interpolation} & 0.137 & 0.112 & 0.110 & 0.102 & 0.100 \\
\hline
\end{tabular}
\end{table}

\begin{figure}[!ht]
\centering
\includegraphics[width=5.1in]{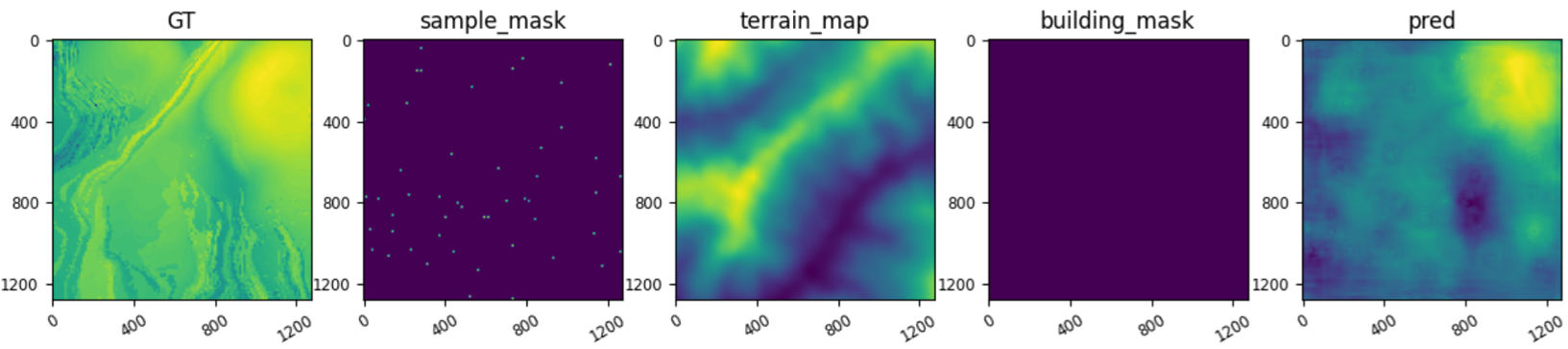}\vspace{-3mm}
\caption{Demonstration for radio map construction in 150 MHz network.}\vspace{-3mm}
\label{150MHzMap}
\end{figure}

\begin{figure}[!ht]
\centering
\includegraphics[width=5.1in]{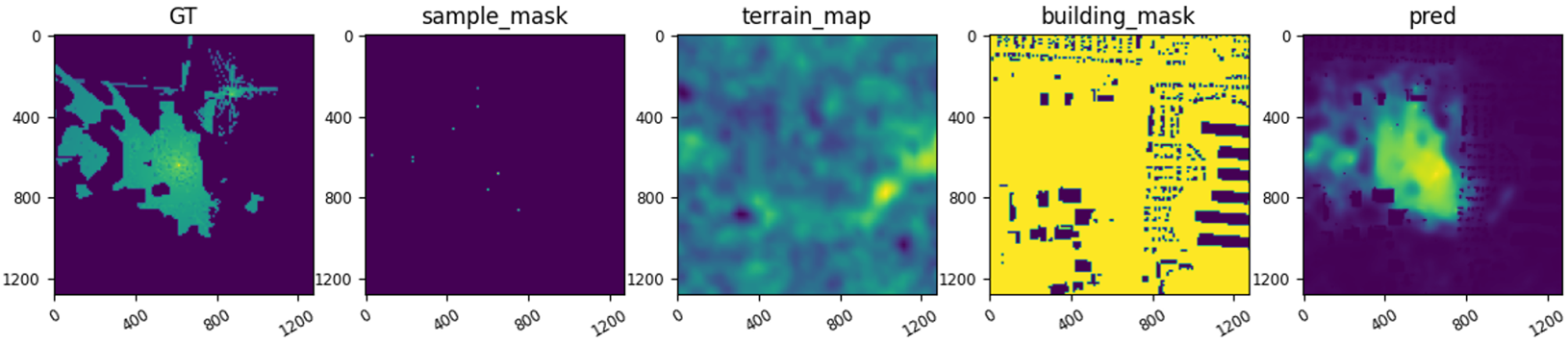}\vspace{-3mm}
\caption{Demonstration for radio map construction in 22 GHz network.}\vspace{-3mm}
\label{22GHzMap}
\end{figure}

Fig.~\ref{150MHzMap} and Fig.~\ref{22GHzMap} provide examples of radio map construction in networks operating at 150 MHz and 22 GHz, respectively. It is evident that terrain and buildings exert a more pronounced shadowing effect on the high-frequency network in Fig.~\ref{22GHzMap}, compared to the low-frequency network in Fig.~\ref{150MHzMap}. The radio map of the high-frequency network exhibits more detailed features due to the influence of subtle terrain variations and small buildings and objects. This observation underscores why constructing radio maps for high-frequency networks is more challenging than for low-frequency networks.

\subsection{Discussions on Generalization}
In this part, we evaluate the generalization of the baseline methods on the \emph{SpectrumNet} dataset across various terrain scenarios and frequency bands.

\subsubsection{Generalization in Different Terrain Scenarios}

To evaluate the generalization of the baseline methods across different terrain scenarios, we train three UNet models and three CBAM models using radio maps of lake, dense urban, and mountainous scenarios, respectively. The performance of these models in constructing radio maps across various terrain scenarios is presented in Table~\ref{LakeExp}, Table~\ref{DenseUrbanExp}, and Table~\ref{MountainousExp}, respectively.

For radio map construction in each terrain scenario, the best-performing model is the one trained on data from the identical terrain scenario. Models trained on data from more complex scenarios, such as dense urban and mountainous, exhibit better generalization than those trained on simpler scenarios, like lakes. To improve the minimum performance of radio map construction across various scenarios, it is essential to include more data from complex scenarios in the training set.

\begin{table}[!ht]
\centering
\caption{Radio map construction models trained with data of lake scenario}\label{LakeExp}
\begin{tabular}{|c|c|c|c|}
\hline
\multicolumn{4}{|c|}{\textbf{SSIM}}\\
\hline
\textbf{Test Scenario} & Lake & Dense urban & Mountainous \\
\hline
\textbf{UNet} & 0.817 & 0.550 & 0.635 \\
\hline
\textbf{CBAM} & 0.831 & 0.510 & 0.665 \\
\hline
\multicolumn{4}{|c|}{\textbf{PSNR}}\\
\hline
\textbf{Test Scenario} & Lake & Dense urban & Mountainous \\
\hline
\textbf{UNet} & 24.508 & 18.947 & 19.390 \\
\hline
\textbf{CBAM} & 25.441 & 18.391 & 20.076 \\
\hline
\multicolumn{4}{|c|}{\textbf{RMSE}}\\
\hline
\textbf{Test Scenario} & Lake & Dense urban & Mountainous \\
\hline
\textbf{UNet} & 0.060 & 0.116 & 0.107 \\
\hline
\textbf{CBAM} & 0.054 & 0.124 & 0.099 \\
\hline
\end{tabular}
\end{table}

\begin{table}[!ht]
\centering
\caption{Radio map construction models trained with data of dense urban scenario}\label{DenseUrbanExp}
\begin{tabular}{|c|c|c|c|}
\hline
\multicolumn{4}{|c|}{\textbf{SSIM}}\\
\hline
\textbf{Test Scenario} & Lake & Dense urban & Mountainous \\
\hline
\textbf{UNet} & 0.817 & 0.638 & 0.672 \\
\hline
\textbf{CBAM} & 0.819 & 0.603 & 0.656 \\
\hline
\multicolumn{4}{|c|}{\textbf{PSNR}}\\
\hline
\textbf{Test Scenario} & Lake & Dense urban & Mountainous \\
\hline
\textbf{UNet} & 23.443 & 19.874 & 19.170 \\
\hline
\textbf{CBAM} & 23.994 & 20.193 & 19.563 \\
\hline
\multicolumn{4}{|c|}{\textbf{RMSE}}\\
\hline
\textbf{Test Scenario} & Lake & Dense urban & Mountainous \\
\hline
\textbf{UNet} & 0.068 & 0.105 & 0.110 \\
\hline
\textbf{CBAM} & 0.064 & 0.101 & 0.105 \\
\hline
\end{tabular}
\end{table}

\begin{table}[!ht]
\centering
\caption{Radio map construction models trained with data of mountainous scenario}\label{MountainousExp}
\begin{tabular}{|c|c|c|c|}
\hline
\multicolumn{4}{|c|}{\textbf{SSIM}}\\
\hline
\textbf{Test Scenario} & Lake & Dense urban & Mountainous \\
\hline
\textbf{UNet} & 0.803 & 0.479 & 0.651 \\
\hline
\textbf{CBAM} & 0.816 & 0.478 & 0.681 \\
\hline
\multicolumn{4}{|c|}{\textbf{PSNR}}\\
\hline
\textbf{Test Scenario} & Lake & Dense urban & Mountainous \\
\hline
\textbf{UNet} & 23.676 & 14.972 & 20.024 \\
\hline
\textbf{CBAM} & 24.428 & 15.400 & 20.896 \\
\hline
\multicolumn{4}{|c|}{\textbf{RMSE}}\\
\hline
\textbf{Test Scenario} & Lake & Dense urban & Mountainous \\
\hline
\textbf{UNet} & 0.066 & 0.185 & 0.100 \\
\hline
\textbf{CBAM} & 0.060 & 0.176 & 0.090 \\
\hline
\end{tabular}
\end{table}

\begin{figure}[!ht]
\centering
\includegraphics[width=5.1in]{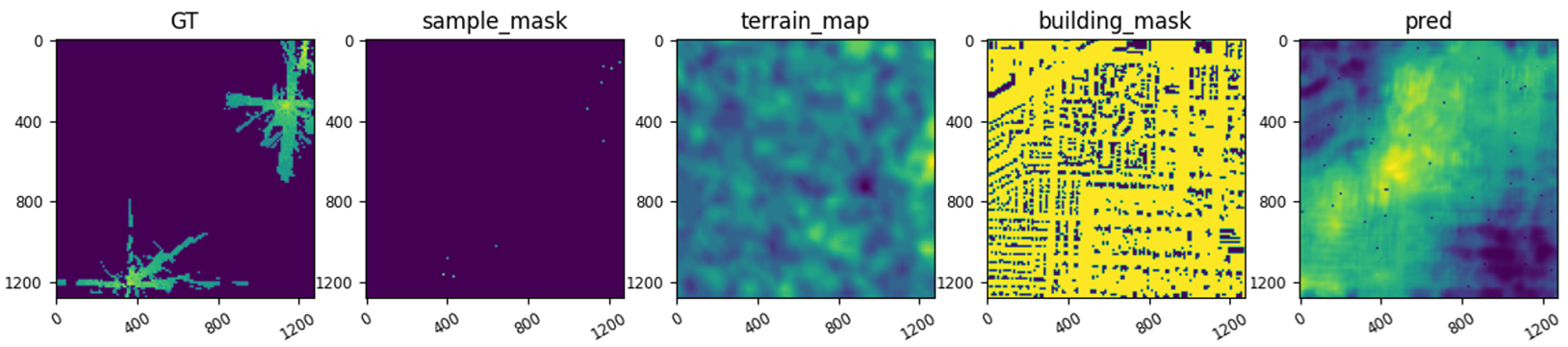}\vspace{-3mm}
\caption{Demonstration for radio map construction in dense urban with model trained with mountainous data.}\vspace{-3mm}
\label{MountainousAdapt}
\end{figure}

\begin{figure}[!ht]
\centering
\includegraphics[width=5.1in]{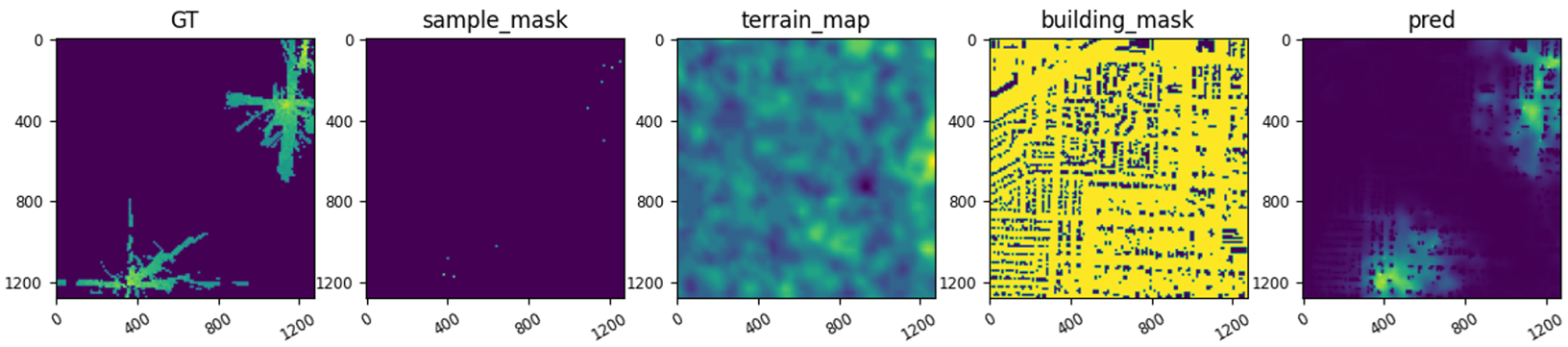}\vspace{-3mm}
\caption{Demonstration for radio map construction in dense urban with model trained with dense urban data.}\vspace{-3mm}
\label{DenseUrbanAdapt}
\end{figure}

Fig.~\ref{MountainousAdapt} and Fig.~\ref{DenseUrbanAdapt} illustrate radio map construction in a dense urban environment using models trained on mountainous and dense urban radio maps, respectively. The radio signal distribution in Fig.~\ref{MountainousAdapt} shows a significant discrepancy from the GT, as the model trained on mountainous maps has limited knowledge of building information. In contrast, the signal distribution in Fig.~\ref{DenseUrbanAdapt} is closer to the GT, although it still lacks precision in capturing detailed features. This experiment highlights the importance of incorporating diverse terrain information into the training set, which is essential for developing generative models with strong generalization capabilities.

\subsubsection{Generalization in Different Frequencies}
We then evaluate generalization across different frequencies by training two UNet models and two CBAM models on radio maps from 150 MHz and 22 GHz networks, respectively. The performance of these models in constructing radio maps for networks across five frequencies is presented in Table~\ref{LowFreqExp} and Table~\ref{HighFreqExp}. Models trained on 22 GHz radio maps outperform those trained on 150 MHz radio maps when applied to high-frequency networks, while the reverse is true for low-frequency networks. This demonstrates the importance of incorporating multiband radio maps into the \emph{SpectrumNet} dataset to achieve high-performance radio map construction across a range of frequencies.

The model trained on radio maps from 150 MHz network performs relatively well in generating radio maps for networks operating at 150 MHz, 1.5 GHz, 1.7 GHz, and 3.5 GHz. However, it struggles to adapt to the 22 GHz network, where radio propagation is more influenced by small scale terrain variations and buildings. In contrast, the model trained on radio maps from 22 GHz network exhibits relatively good performance across all five frequencies, as it learns to account for both large scale and small scale terrain variations and building impacts on radio propagation. Thus, incorporating data from high-frequency networks is crucial for training a model with strong generalization across different frequencies.

\begin{table}[!ht]
\centering
\caption{Radio map construction models trained with data of 150 MHz}\label{LowFreqExp}
\begin{tabular}{|c|c|c|c|c|c|}
\hline
\multicolumn{6}{|c|}{\textbf{SSIM}}\\
\hline
\textbf{Test Scenario} & 150 MHz & 1.5 GHz & 1.7 GHz & 3.5 GHz & 22 GHz \\
\hline
\textbf{UNet} & 0.719 & 0.641 & 0.637 & 0.595 & 0.461 \\
\hline
\textbf{CBAM} & 0.740 & 0.665 & 0.665 & 0.631 & 0.458 \\
\hline
\multicolumn{6}{|c|}{\textbf{PSNR}}\\
\hline
\textbf{Test Scenario} & 150 MHz & 1.5 GHz & 1.7 GHz & 3.5 GHz & 22 GHz \\
\hline
\textbf{UNet} & 19.073 & 19.700 & 19.795 & 19.476 & 18.908 \\
\hline
\textbf{CBAM} & 19.522 & 20.475 & 20.656 & 20.944 & 20.495 \\
\hline
\multicolumn{6}{|c|}{\textbf{RMSE}}\\
\hline
\textbf{Test Scenario} & 150 MHz & 1.5 GHz & 1.7 GHz & 3.5 GHz & 22 GHz \\
\hline
\textbf{UNet} & 0.113 & 0.105 & 0.104 & 0.108 & 0.115 \\
\hline
\textbf{CBAM} & 0.107 & 0.096 & 0.094 & 0.091 & 0.096 \\
\hline
\end{tabular}
\end{table}

\begin{table}[!ht]
\centering
\caption{Radio map construction models trained with data of 22 GHz}\label{HighFreqExp}
\begin{tabular}{|c|c|c|c|c|c|}
\hline
\multicolumn{6}{|c|}{\textbf{SSIM}}\\
\hline
\textbf{Test Scenario} & 150 MHz & 1.5 GHz & 1.7 GHz & 3.5 GHz & 22 GHz \\
\hline
\textbf{UNet} & 0.645 & 0.666 & 0.668 & 0.658 & 0.629 \\
\hline
\textbf{CBAM} & 0.646 & 0.677 & 0.678 & 0.666 & 0.637 \\
\hline
\multicolumn{6}{|c|}{\textbf{PSNR}}\\
\hline
\textbf{Test Scenario} & 150 MHz & 1.5 GHz & 1.7 GHz & 3.5 GHz & 22 GHz \\
\hline
\textbf{UNet} & 15.297 & 19.422 & 19.703 & 20.780 & 22.260 \\
\hline
\textbf{CBAM} & 16.537 & 19.955 & 20.224 & 21.205 & 22.529 \\
\hline
\multicolumn{6}{|c|}{\textbf{RMSE}}\\
\hline
\textbf{Test Scenario} & 150 MHz & 1.5 GHz & 1.7 GHz & 3.5 GHz & 22 GHz \\
\hline
\textbf{UNet} & 0.174 & 0.108 & 0.105 & 0.093 & 0.078 \\
\hline
\textbf{CBAM} & 0.151 & 0.102 & 0.099 & 0.088 & 0.076 \\
\hline
\end{tabular}
\end{table}

\begin{figure}[!ht]
\centering
\includegraphics[width=5.1in]{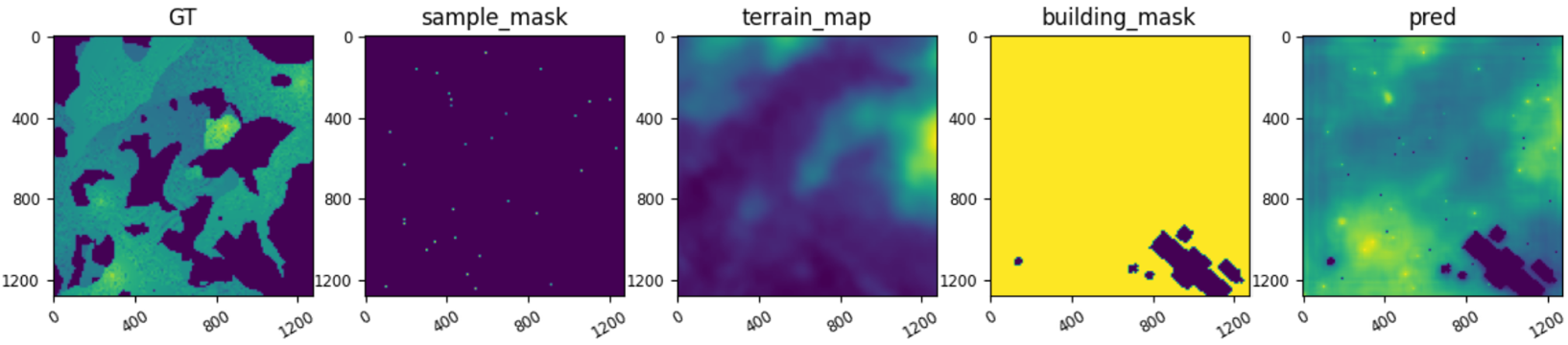}\vspace{-3mm}
\caption{Demonstration for radio map construction in 22 GHz network with model trained by data of 150 MHz network.}\vspace{-3mm}
\label{150MHzAdapt}
\end{figure}

\begin{figure}[!ht]
\centering
\includegraphics[width=5.1in]{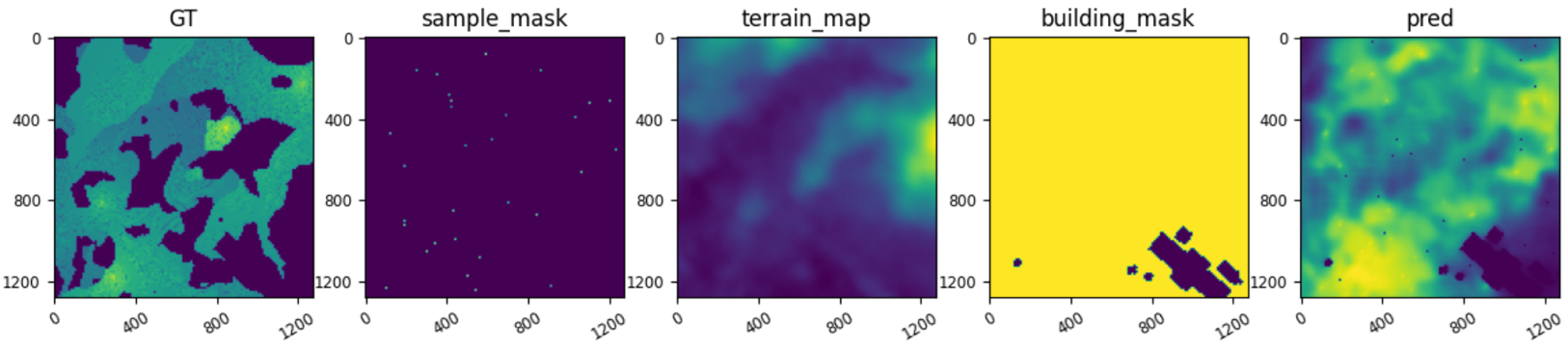}\vspace{-3mm}
\caption{Demonstration for radio map construction in 22 GHz network with model trained by data of 22 GHz network.}\vspace{-3mm}
\label{22GHzAdapt}
\end{figure}

A comparison of radio map construction for 22 GHz network using models trained on 150 MHz and 22 GHz radio maps is shown in Fig.~\ref{150MHzAdapt} and Fig.~\ref{22GHzAdapt}, respectively. In Fig.~\ref{150MHzAdapt}, the model trained on 150 MHz radio maps primarily captures large scale fading effects, with insufficient attention to small scale terrain impact. Consequently, the 22 GHz radio map constructed by this model shows significant performance gaps, particularly in terrain effects and detailed features. In contrast, the model trained on 22 GHz radio maps, as depicted in Fig.~\ref{22GHzAdapt}, provides clear detail descriptions and effectively captures the terrain's impact on radio propagation.

\section{Discussion and Future Works for Dataset Expansion}\label{DiscussionSec}
In this section, we discuss future works on expanding the \emph{SpectrumNet} dataset, and the extended applications of generative AI with the \emph{SpectrumNet} dataset.

\subsection{Dataset Expansion}

\subsubsection{Spacial Expansion}
The projection area of the samples in the current version of the \emph{SpectrumNet} dataset is $1.28 km \times 1.28 km$, which is insufficient to accommodate the radio maps of networks with large cell radius, e.g. multimedia broadcast multicast service networks, where the cell radius can exceed $2 km$~\cite{TANI2011}. In future versions of the \emph{SpectrumNet} dataset, the sample area can be expanded to $5km \times 5km$ while maintaining the same resolution. This expansion will allow generative models for radio map construction to learn the spatial correlation knowledge of networks with large cell radius.

\subsubsection{Full Band Dataset Construction}
Although the current version of the \emph{SpectrumNet} dataset contains multiple working frequencies, the frequency domain resolution is relatively low compared to the granularity of spectrum resource allocation. Future versions of the \emph{SpectrumNet} dataset will provide higher resolution in the frequency domain, encompassing all commonly used bands for wireless networks.

\subsubsection{Time Domain Dataset Construction}
Radio map data exhibits time domain correlation in terms of transmitter location, transmission power, climate parameters, etc. However, all existing radio map datasets omit this correlation. In future versions of the \emph{SpectrumNet} dataset, multiple radio maps in consecutive time slots will be included. This time domain correlation can be utilized to train generative models for predicting the track of moving transmitters, the power variation of each transmitter, and the change in climate parameters. Based on these predictive models, the radio map for upcoming time slots can be forecasted, enabling applications such as channel prediction and other radio map prediction-based functionalities.

\subsubsection{Directional Radio Propagation}
In the current version of the \emph{SpectrumNet} dataset, the antennas of the transmitters are set to be omnidirectional. However, with the development of multi-antenna systems, directional beamforming has become more commonly applied~\cite{BHMDR2017}. Future versions of the \emph{SpectrumNet} dataset will include cases reflecting the radio propagation properties of anisotropic RSS distribution caused by directional transmissions.

\subsubsection{Dataset Calibration}
Radio map datasets such as those in~\cite{RYKC2021, LZLLXXMZQX2024, JCD2024} consist of radio maps simulated with ray-tracing-based software, and the accuracy of these simulation platforms has not been verified. In future work, we plan to supplement the radio map data with hardware experiments. We will analyze the similarity between the data collected in realistic scenarios and the simulation data to measure the accuracy of the radio maps. Furthermore, the simulation parameters, such as reflection and refraction coefficients of different materials, can be further calibrated based on the collected data.

\subsection{Extended Studies on Generative Models for Radio Map Construction}
With the open-source \emph{SpectrumNet} dataset, further applications on data-driven radio map construction can be studied. In this part, we discuss several research directions on generative model for radio map construction using the \emph{SpectrumNet} dataset.

\subsubsection{Generative Model Enhancement}
When comparing the performance of generative models for radio map construction in Section~\ref{ApplicationSec} with the works on other datasets~\cite{RYKC2021,LZLLXXMZQX2024,JCD2024}, we find that the baseline models cannot provide high accuracy radio map construction for multiband 3D scenario with terrain information. Therefore, designing training methods for generative models with the \emph{SpectrumNet} dataset is necessary and can be explored in the following ways:

\begin{enumerate}
\item{\textbf{Model scale extension}:} As mentioned above, the parameters and dimensions of the \emph{SpectrumNet} dataset are much larger than other existing radio map datasets~\cite{RYKC2021,LZLLXXMZQX2024,JCD2024}. It is very difficult for models of comparable scale to those in~\cite{RYKC2021,LZLLXXMZQX2024,JCD2024} to learn the extensive knowledge embedded within the \emph{SpectrumNet} dataset. In order to train a high accuracy generative model for radio map construction based on the \emph{SpectrumNet}, the model scale should be enhanced.
\item{\textbf{Model framework design}:} Most of the current works on radio map construction utilize generative model frameworks from other visual tasks, such as UNet~\cite{RFB2015}, MAE~\cite{HCXLDG2022}, etc. However, these frameworks may struggle with radio map construction, where the model input is extremely sparse (less than 1\%~\cite{SHVR2023}) and of pixel level. As a result, novel framework designs for radio map construction should be further studied.
\item{\textbf{Channel model assisted model enhancement}:} It is very challenging and costly to train generative model for radio map construction barely based on datasets with a sophisticated knowledge, such as the \emph{SpectrumNet}. Leveraging the channel model of wireless networks can simplify the tasks for generative models in radio map construction. For example, a channel model based radio map construction can be added as a preprocessing step for input data. Additionally, it can be incorporated into the loss function for model training.
\end{enumerate}

\subsubsection{Cross Domain Radio Map Construction}
Current studies on generative models for radio map construction typically focus on one or two specific domains due to the limited dimensions in existing datasets. The \emph{SpectrumNet} is the first dataset that includes diversities in space, frequency, terrain, and weather domains.\footnote{It will also contain time domain in the future.} With the radio maps in the \emph{SpectrumNet} dataset, cross domain generative models for radio map construction can be explored as a new research direction. For example, once a space-frequency cross domain model is trained, it becomes possible to estimate the channel of an aerial mmWave network by sampling the signals of a C-band terrestrial network. Cross domain radio map construction is especially valuable for estimating the network conditions of areas and spectrums that cannot be measured directly.

\section{Conclusions}\label{Conclusions}
In this paper, we have introduced an open-source radio map dataset \emph{SpectrumNet}, which is a multiband 3D radio map dataset with the consideration of terrain and climate information. We have described the process of constructing the radio maps incorporating both terrain and climate factors, and have illustrated the radio maps in both the frequency and spatial 3D domains. Experiments on training generative models for radio map construction have been conducted using the \emph{SpectrumNet} dataset, demonstrating the necessity of this dataset for model generalization. We have also discussed future works on further expanding the \emph{SpectrumNet} dataset and extending studies on generative models for radio map construction. A few conclusions are summarised as follows.
\begin{enumerate}
\item \emph{SpectrumNet} is the largest radio map dataset in terms of dimensions and scale. It is a spacial 3D radio map dataset that includes 5 frequency bands, 11 terrain scenarios, and 3 climate scenarios. The generation of each sample in the \emph{SpectrumNet} has fully considered the physical properties of radio propagation, including signal reflection and refraction on various terrain types, climates, and object materials.
\item Three different methods for radio map construction, i.e., UNet, CBAM, and interpolation algorithm, have been evaluated based on the \emph{SpectrumNet} dataset. Experimental results have shown that baseline models trained on data of dense urban scenario, terrestrial networks, and high frequency networks have strong generalization when applied to other terrain scenarios, heights, and frequencies. It has also been proved that the existed datasets of single band, height, and terrain scenario, are insufficient for training radio map construction models with generalization capability.
\item Future works on \emph{SpectrumNet} have been discussed in this paper, including the enhancements of the \emph{SpectrumNet} dataset through domain expansion and data calibration with real-world data collection. Additionally, the \emph{SpectrumNet} dataseet supports new research directions for generative models in radio map construction, such as model framework design and cross domain model training.
\end{enumerate}

\begin{appendices}
\section{File Introduction to the \emph{SpectrumNet} Dataset}\label{Appendix1}
In the following, we introduce the files of the \emph{SpectrumNet} dataset. The properties of each radio map are marked in the file names, with the naming conventions outlined in Table~\ref{FileName}.
 \begin{table}[!ht]
\centering
\caption{File Name Explanation}\label{FileName}
\begin{tabular}{|c|c|c|}
\hline
\textbf{Field name} & \textbf{Property} & \textbf{Number meaning} \\
\hline
T + 2 digits & Geographic type & Digits from 01 to 11, for different types of geographical scenarios, as shown in Table~\ref{GeoType}. \\
\hline
C + 1 digit & Climate type & Digit from 0 to 2, for different types of climate, as shown in Table~\ref{CliType}. \\
\hline
D + 4 digits & Map number & Digits start from 0000, for different maps. \\
\hline
n + 2 digits & Sampling number & Digits from 00, for different transmitter samplings in a specific map. \\
\hline
f + 2 digits & Frequency & Digits from 00 to 04, for different frequencies, as shown in Table~\ref{FreNum}. \\
\hline
z + 2 digits & Height & Digit from 00 to 02, for different heights of radio map, as shown in Table~\ref{HeiType}.  \\
\hline
\end{tabular}
\end{table}

\begin{table}[!ht]
\centering
\caption{Geographic Type}\label{GeoType}
\begin{tabular}{|c|c|c|c|c|c|c|}
\hline
\textbf{Digits} & T01 & T02 & T03 & T04 & T05 & T06 \\
\hline
\textbf{Meaning} & Grassland & Island & Ocean & Lake & Suburban & Dense urban \\
\hline
\textbf{Digits} & T07 & T08 & T09 & T10 & T11 &\\
\hline
\textbf{Meaning} & Rural & Ordinary urban & Desert & Mountainous & Forest &\\
\hline
\end{tabular}
\end{table}

\begin{table}[!ht]
\centering
\caption{Climate Type}\label{CliType}
\begin{tabular}{|c|c|c|c|}
\hline
\textbf{Digits} & C0 & C1 & C2 \\
\hline
\textbf{Meaning} & Tropics & Subtropics & Temperature zone\\
\hline
\end{tabular}
\end{table}

\begin{table}[!ht]
\centering
\caption{Frequency Number}\label{FreNum}
\begin{tabular}{|c|c|c|c|c|c|}
\hline
\textbf{Digits} & f00 & f01 & f02 & f03 & f04  \\
\hline
\textbf{Meaning} & 150 MHz & 1.5 GHz & 1.7 GHz & 3.5 GHz & 22 GHz\\
\hline
\end{tabular}
\end{table}

\begin{table}[!ht]
\centering
\caption{Height Number}\label{HeiType}
\begin{tabular}{|c|c|c|c|}
\hline
\textbf{Digits} & z00 & z01 & z02 \\
\hline
\textbf{Meaning} & 1.5 m & 30 m & 200 m\\
\hline
\end{tabular}
\end{table}

Three different types of files are included in the \emph{SpectrumNet} dataset, i.e., npz, jpg, and png files. Each sample from one area contains 1 npz file, 5 jpg files, and 15 png files. The contents of these files are explained as below.
\begin{enumerate}[(1)]
\item \textbf{npz:} Files of npz format contains the terrain and building information of a specific region. The naming rule of npz file is ``T[xx]C[x]D[xxxx]\underline{ }n[xx]\underline{ }bdtr". The meaning of the numbers in ``[]" can be found in Table~\ref{FileName}.
\item \textbf{jpg:} A file of jpg format is a visualization image of radio map and terrain information for one region at one frequency. The naming rule of jpg file is ``T[xx]C[x]D[xxxx]\underline{ }n[xx]\underline{ }f[xx]\underline{ }visual". The meaning of the numbers in ``[]" can be found in Table~\ref{FileName}.
\item \textbf{png:} A file of png format is a 2D radio map for one region at one frequency and one height. The naming rule of png file is ``T[xx]C[x]D[xxxx]\underline{ }n[xx]\underline{ }f[xx]\underline{ }ss\underline{ }z[xx]". The meaning of the numbers in ``[]" can be found in Table~\ref{FileName}.
\end{enumerate}
\end{appendices}

\end{document}